\documentclass{article}
\usepackage[utf8]{inputenc}
\usepackage{spconf,amsmath,graphicx}
\usepackage{booktabs}
\usepackage{multirow}
\usepackage{multicol}
\usepackage{bm}
\usepackage{subcaption}
\PassOptionsToPackage{hyphens}{url}\usepackage[colorlinks=false]{hyperref}

    \def\itemautorefname~#1\null{#1\null}
\usepackage{todonotes}
\usepackage{tikz}
\usetikzlibrary{shapes, arrows, positioning, shapes.arrows, decorations.text, external}
\usepackage{xcolor}

\DeclareMathOperator{\diag}{diag}

\DeclareMathOperator*{\argmax}{argmax}

\title{Bringing in the outliers: A sparse subspace clustering approach to learn a dictionary of mouse ultrasonic vocalizations\thanks{Paper accepted in ICASSP 2020.}}

\name{Jiaxi Wang$^{\star}$\thanks{Jiaxi Wang's work was performed during an internship at USC SAIL. He is now with the Department of Electrical Engineering, Tsinghua University.} \qquad Karel Mundnich$^{\star}$ \qquad Allison T. Knoll $^{\dagger}$ \qquad Pat Levitt$^{\dagger}$ \qquad Shrikanth Narayanan$^{\star}$}
\address{\small{$^{\star}$ Signal Analysis and Interpretation Lab, University of Southern California, Los Angeles, CA 90089, USA} \\ \small{$^{\dagger}$ Department of Pediatrics and Program in Developmental Neuroscience and Neurogenetics, The Saban Research Institute,}\\\small{Children's Hospital Los Angeles, Keck School of Medicine, University of Southern California, Los Angeles, CA 90089, USA}}
\begin{document}
\ninept
%
\maketitle
\begin{abstract}
Mice vocalize in the ultrasonic range during social interactions. These vocalizations are used in neuroscience and clinical studies to tap into complex behaviors and states. The analysis of these ultrasonic vocalizations (USVs) has been traditionally a manual process, which is prone to errors and human bias, and is not scalable to large scale analysis. We propose a new method to automatically create a dictionary of USVs based on a two-step spectral clustering approach, where we split the set of USVs into inlier and outlier data sets. This approach is motivated by the known degrading performance of sparse subspace clustering with outliers. We apply spectral clustering to the inlier data set and later find the clusters for the outliers. We propose quantitative and qualitative performance measures to evaluate our method in this setting, where there is no ground truth. Our approach outperforms two baselines based on k-means and spectral clustering in all of the proposed performance measures, showing greater distances between clusters and more variability between clusters.
\end{abstract}

\section{Introduction}
\label{sec:intro}
Mice vocalize in the ultrasonic range between 30 and 150kHz \cite{sewell1968ultrasound}. These vocalizations are primarily observed during social interactions and their properties vary across different social contexts \cite{Portfors2014The, Chabout2014Male, Cao2014A}. Given the social nature of this behavior, mouse ultrasonic vocalizations (USVs) have become of special interest in biomedical research to obtain information about complex social communication behaviors, especially in genetic models of neurodevelopmental disorders \cite{crawley2012translational} by using vocal communication as a proxy \cite{fischer2011ultrasonic, W2013Affective} to study different behavioral patterns and states.

Generating a dictionary (inventory) of USVs is a critical step in the behavioral analysis of social communication and interaction. The study of vocalizations has traditionally been performed in a manual process, where human annotators annotate and cluster USVs into a small number of groups (4--10) according to subjective rules, including their shape in the frequency domain and their duration \cite{scattoni2011unusual}. However, this is a time-consuming process that is prone to annotation errors and annotator biases, both of which potentially affect subsequent analyses, and which may not necessarily reflect the structure of the data. Therefore, the problem of finding a dictionary of vocalizations in animals using automated tools has started to attract interest in recent years \cite{arriaga2012mice, burkett2015voice, Van2017MUPET, coffey2019deepsqueak}.

\begin{figure}[t]
  \centering
  \includegraphics[width=\columnwidth]{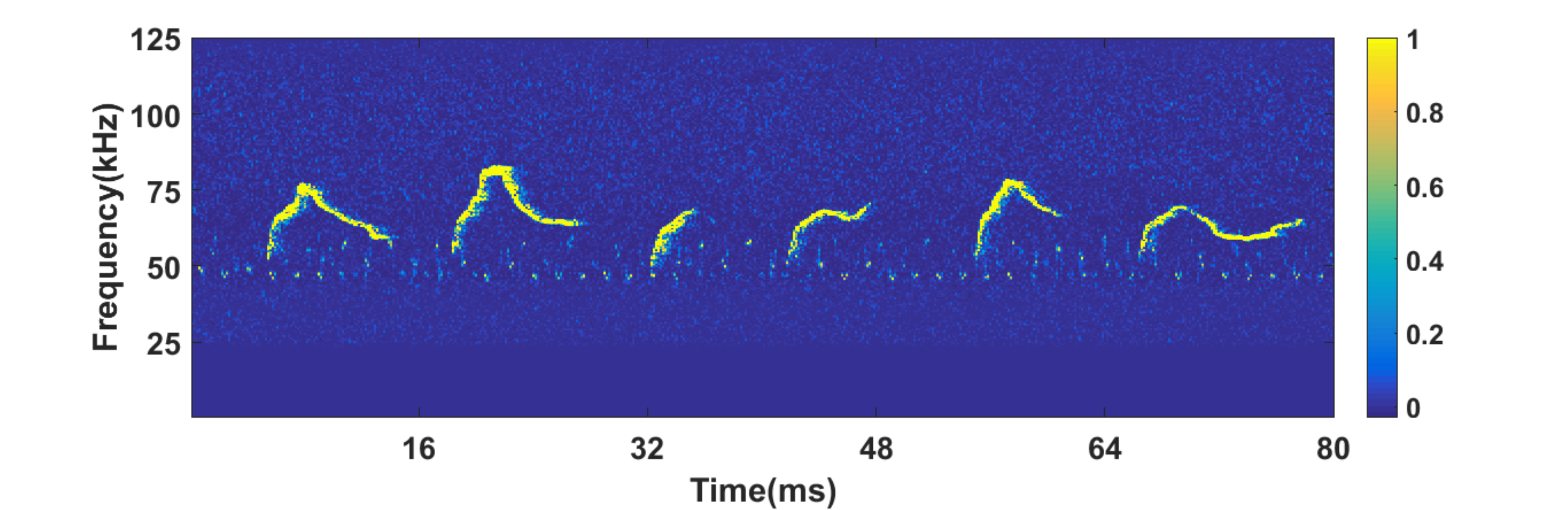}
  \caption{A spectrogram depicting a sequence of mouse ultrasonic vocalizations (USVs). Brighter colors represent higher energy in a given frequency band.}
\label{fig:usv_examples}
\vspace{-0.25cm}
\end{figure}

To analyze the sounds emitted by mice, it is common to analyze the spectrograms of the ultrasonic audio signals. An example of these vocalizations in the frequency domain is shown in \autoref{fig:usv_examples}, where different shapes can be identified in time in the spectrogram of the ultrasonic audio signal. From \autoref{fig:usv_examples}, we can see that the clustering of USVs into groups is a non-trivial problem, due to the lack of ground truth (the true number of different vocalization types is unknown), no intuitive distance or similarity between the USVs to aid the comparison between them, and the difficulty involved in creating rules that generalize over different sets of USVs (for example, generated by different strains of mice).

\subsection{Related work}
Different approaches have been proposed in the literature to analyze USVs in an automated fashion. A common first step is to detect the USVs in an audio signal. Several automated tools have been proposed \cite{burkett2015voice, Van2017MUPET, coffey2019deepsqueak, Reno2013Automating, Barker2014Automated}, all of which extract these vocalizations from the spectrograms of the audio signal through different methods. For example, MUPET \cite{Van2017MUPET} detects the vocalizations based on the frame energy after a noise-reduction step, in a similar fashion to voice activity detection (VAD) in quiet environments. DeepSqueak \cite{coffey2019deepsqueak} takes a computer vision approach, where a Faster-RCNN \cite{ren2015faster} is trained to detect the vocalizations in spectrograms as in object detection tasks.

The clustering approach to learn a dictionary of USVs was proposed in MUPET \cite{Van2017MUPET}, and later employed in DeepSqueak \cite{coffey2019deepsqueak}. Both methods use k-means at their core: \cite{Van2017MUPET} uses features obtained from the audio signal through a gammatone filterbank to perform clustering, while \cite{coffey2019deepsqueak} uses intuitive features such as shape, frequency, and duration in time to cluster the USVs. In MUPET, the number of clusters is user-defined, and different clusters have overlapping features (leading to clusters that are noisy). A different approach is taken by MSA \cite{arriaga2012mice}, where a small number of clusters or categories is used (often only 4), leading to high variability in the shapes of USVs within clusters: upward shapes that are both long and short, as well as complex shapes with different feature values.

\subsection{Contributions}
In this work, we propose a novel approach to cluster USVs based on the hypothesis that the information in the frequency domain lies in low-dimensional subspaces. We are motivated by the notion of distances between subspaces proposed in \cite{Soltanolkotabi2013Robust}, and use these distances to perform sparse subspace clustering (SSC) over sets of segmented USVs. Since sparse subspace clustering is sensitive to outliers, we also propose a two-step approach to sparse subspace clustering, where we first cluster an inlier set of USVs and then add outliers back into our clusters. Finally, we propose a new way to evaluate the quality of the clusters for this particular task.

\section{Background: Subspace Clustering}
\label{sec:subspace_clustering}
Subspace clustering aims to find subspaces in which the data samples lie by posing the clustering problem as a regression problem.
In particular, sparse subspace clustering (SSC) \cite{ssc1, ssc2} is a popular version of subspace clustering methods due to its good performance in different application areas, as well as theoretical guarantees \cite{Soltanolkotabi2013Robust, Soltanolkotabi2011A}.
For an in-depth overview of the methods, we refer the readers to \cite{Vidal2012A}.

The basic idea behind SSC is that data points that lie in a linear subspace can be linearly represented by other data points located in the same subspace. The coefficients that represent these subspaces, which can be found through the LASSO \cite{Tibshirani1996Regression}, reflect the similarity between different data points and can be used to build an affinity matrix. Then, spectral clustering \cite{ng2002spectral, von2007tutorial} is applied to this affinity matrix to get different clusters. According to \cite{Soltanolkotabi2013Robust, Soltanolkotabi2011A}, the success of SSC depends mainly on three properties of the data: (1) low affinity between subspaces, (2) enough samples for a certain subspace, and (3) the data not having too many outliers (where an outlier is defined as a data point that does not lie in any of the subspaces).

Formally, sparse subspace clustering can be posed as an $\ell_1$ minimization problem over the subspace coefficients:
\begin{equation}\label{eq:l1-norm-exact}
    \min_{\bm{y}}\|\bm{y}\|_{1} \quad s.t.\quad \bm{s} = \bm{S}^{*}\bm{y},
\end{equation}
where $\bm{y}$ contains the coefficients of the subspace and $\bm{S}^{*}$ contains all (vectorized) USVs in its column, except for USV $\bm{s}$.

An equivalent formulation uses LASSO instead:
\begin{equation}\label{eq:lasso-formulation}
    \min_{\bm{y}} \frac{1}{2} \|\bm{s} - \bm{S}^{*}\bm{y}\|_2^2 + \lambda \|\bm{y}\|_1.
\end{equation}
We can rewrite \autoref{eq:lasso-formulation} in matrix form to include all USVs $\bm{s}_i, i\in\{1,\ldots,N\}$:
\begin{equation}\label{eq:lasso-formulation-matrix-form}
    \min_{\bm{y}} \frac{1}{2} \|\bm{S} - \bm{S}\bm{Y}\|_2^2 + \lambda \|\bm{Y}\|_1 \quad s.t. \quad \diag(\bm{Y}) = 0,
\end{equation}
where $\diag(\bm{Y})$ is a vector with the diagonal elements of $\bm{Y}$. Each column of $\bm{S}$ contains a vocalization and each column of $\bm{Y}$ contains the coefficients for the different USVs.

From $\bm{Y}$, the similarity matrix $\bm{A}$ is computed by:
\begin{equation}\label{eq:similarity-matrix}
    \bm{A} = |\bm{Y}|+|\bm{Y}|^{T}.
\end{equation}
The elements in $\bm{A}$ are the subspace similarities. To find the clusters, we apply spectral clustering (with random walk Laplacian) \cite{von2007tutorial, shi2000normalized} to $\bm{A}$.

\section{Proposed Method}\label{sec:proposed-method}
\begin{figure}[t]

\centering
\includegraphics[scale=0.75]{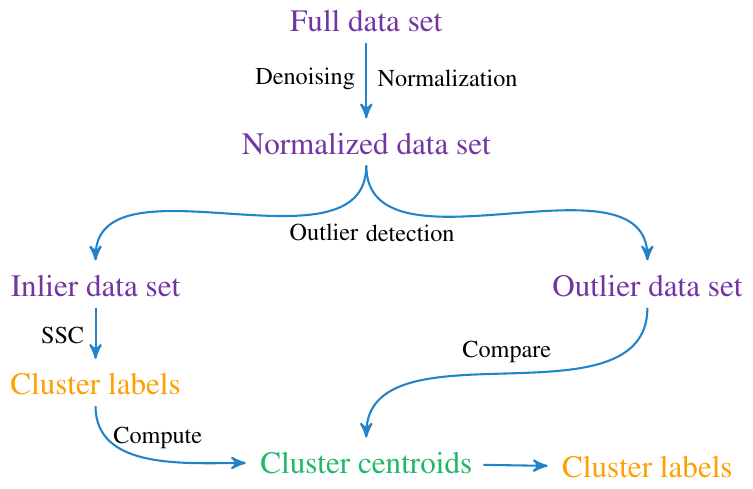}
\caption{Summary of the proposed method.}
\label{fig:algorithm_summary}
\vspace{-0.25cm}
\end{figure}

Our proposed method is a hybrid approach between sparse subspace clustering and more traditional clustering techniques. We are motivated by the fact that sparse subspace clustering does not perform well in the case of many outliers, which are expected to exist in spectrally diverse and non-stereotyped mouse USV data sets. Therefore, we propose a simple approach for outlier detection, perform sparse subspace clustering on the inlier data set, and then add the outliers to our clusters. The steps for the method are as follows: (1) detection of USVs using MUPET, (2) pre-processing of the segmented USVs, (3) dividing our data set into inliers and outliers, (4) performing subspace clustering on the inlier data set, and (5) assigning a cluster to each USV in the outlier set. These steps are summarized in \autoref{fig:algorithm_summary}.

\subsection{Preprocessing}
\subsubsection{Segmentation, denoising and normalization}
To detect and segment USVs from audio files, we use MUPET. After segmentation, the vocalizations may have different time durations or be located in distinct frequency bands. Therefore, we normalize the vocalizations to a same size $F \times T$ (note that we depart from MUPET in this step).
Therefore, we represent each USV by employing the regions in the frequency domain whose energy is greater than the average energy, by clipping the areas with energy below the average energy.
We note that the segmented USVs contain energy in different frequency bands, and are of different time-lengths.
Therefore, we use MATLAB's \texttt{imresize} function (which implements a bi-cubic interpolation) to normalize the segmented USVs into a same size, after which the USVs with similar shapes are frequency-aligned and have the same length in time. After the normalization, we vectorize the USV matrices.

\subsubsection{Outlier detection}
Since the performance of sparse subspace clustering highly degrades in the case of a large amount of outliers \cite{Soltanolkotabi2013Robust}, we first remove the outliers and perform sparse subspace clustering on the resulting inlier data set.

Intuitively, an outlier is an USV that differs from any of all other USVs. Let $\bm{s}_{i}$ denote the vector of the $i^\text{th}$ vocalization, where $i\in\{1,\ldots,N\}$. We define a vocalization $\bm{s}$ as an outlier if:
\begin{equation}\label{eq:threshold}
    \max_{\bm{s} \neq \bm{s}'} \; \cos(\bm{s}, \bm{s}') < \tau,
\end{equation}
where $\tau$ is a threshold to be chosen and $\cos(\cdot,\cdot)$ is the \textit{cosine similarity} (CS) between vectors:
\begin{equation}
    \cos(\bm{s}, \bm{s}') = \frac{\bm{s}\cdot\bm{s}'}{\|\bm{s}\|\|\bm{s}'\|}.
\end{equation}

\subsection{Step one: clustering inliers}
We perform sparse subspace clustering as described in \autoref{sec:subspace_clustering} in the inlier data set. In this case, $\bm{S}$ is the matrix of (vectorized) USVs and $\bm{Y}$ is the matrix of subspace coefficients.

\subsection{Step two: clustering outliers}
After assigning clusters to the inlier data set, we assign clusters to each one of the USVs in the outlier data set.
There are mainly two ways of assignment: (1) we can cluster the outliers into existing categories, and (2) we can create new clusters for outliers.
Here, we simply cluster the outliers into existing categories, and leave the problem of assigning novel clusters for future work.
To perform cluster assignments, we first define the concept of \textit{centroids} in this setting as the mean of the USVs inside a cluster. Assume $\bm{c}_k$ is the centroid of cluster $k$ we get from sparse subspace clustering, $k=1,2,...,K$ where $K$ is the number of clusters. We assign an outlier $\bm{s}_{out}$ into the most similar cluster $k$ using:
\begin{equation}
    k = \argmax_{j} \cos(\bm{s}_{out}, \bm{c}_j).
\end{equation}

\section{Experiments}\label{sec:experiments}
We now describe our experiments. All the code is available online\footnote{\url{www.github.com/usc-sail/mupet-subspace-clustering}}.

\subsection{Data}
The data set\footnote{\url{https://github.com/mvansegbroeck/mupet/wiki/MUPET-wiki} (sample files for DBA and C57).} we use contains 40 records sampled at 250kHz emitted by both DBA/2J (DBA) and C57Bl/6J (C57) mouse strains. After vocalization detection using MUPET \cite{Van2017MUPET}, we used approximately 9000 vocalizations per strain. For details on how the data was collected, we refer readers to \cite{knoll2018quantitative}.

\subsection{Parameters}
We set $F = T = 64$, such that each vector $\bm{s}$ is of length $64\times 64$. We choose the outlier threshold $\tau=0.8$ for DBA and $\tau=0.7$ for C57 (\autoref{eq:threshold}). According to \cite{Soltanolkotabi2013Robust}, the selection of $\lambda$ depends on the dimension of the subspaces, which remains unknown in our case. In practice, we find the dimension of the subspaces empirically by trial and error, and settle for $\lambda = 0.3$ for both data sets, which contains moderately sparse represented coefficients. We also test the different clustering methods with $K = 20, 40$, and $60$ clusters.

Given that small perturbations in the adjacency matrix $\bm{A}$ may influence the eigenvectors of the Laplacian matrix computed during spectral clustering (\autoref{eq:similarity-matrix}), negatively affecting the results of the clustering performance, we set the elements in $\bm{Y}$ smaller than $0.001$ to $0$. We regard these values of $\bm{Y}$ as noise.

\subsection{Baselines and Methods}
We compare our proposed approach with two different baselines. The first one is the k-means approach used by MUPET. The second baseline uses spectral clustering by computing a similarity matrix based on the cosine similarity (CS), where the similarity is defined as:
\begin{equation}
    \bm{A}_{ij} = \cos(\bm{s}_i, \bm{s}_j)
\end{equation}
between USVs $\bm{s}_i$ and $\bm{s}_j$. We also use two approaches to solve the SSC problem: Orthogonal Matching Pursuit (OMP-SSC) \cite{you2016scalable} to solve the matrix form of \autoref{eq:l1-norm-exact} and the LASSO formulation in \autoref{eq:lasso-formulation-matrix-form} (LASSO-SSC).

\begin{table}[t]
    \centering
    \caption{Results for the clustering methods. CS + SC: spectral clustering using cosine similarity, LASSO-SSC: LASSO-based subspace clustering, OMP-SSC: OMP-based subspace clustering.}
    \vspace{-0.2cm}
    \begin{tabular}{lclcc}
        \toprule
        \textbf{Strain} & $\bm{K}$ & \textbf{Method} & \multicolumn{2}{c}{\textbf{Inliers only / With outliers}} \\
        \cmidrule{4-5}
                        &  &  & $\bar{d}_{\cos}(\bm{C})$ & $\sigma(d_{\cos}(\bm{C}))$ \\
        \midrule
        \multirow{13}{*}{\textbf{DBA}} & \multirow{4}{*}{20} & k-means   & 0.374 / 0.372 & 0.215 / 0.209 \\
                                      &                      & CS + SC   & 0.428 / 0.418 & 0.209 / 0.200 \\
                                      &                      & LASSO-SSC & \textbf{0.572} / \textbf{0.463} & \textbf{0.168} / \textbf{0.164} \\
                                      &
                              & OMP-SSC & 0.296 / 0.315 &  0.242 / 0.229\\
        \cmidrule{2-5}
                                      & \multirow{4}{*}{40} & k-means   & 0.383 / 0.380 & 0.223 / 0.216 \\
                                      &                     & CS + SC   & 0.448 / 0.439 & 0.215 / 0.206 \\
                                      &                     & LASSO-SSC & \textbf{0.529} / \textbf{0.484} & \textbf{0.178} / \textbf{0.174} \\
                                      &
                              & OMP-SSC & 0.212 / 0.225 & 0.271 / 0.257 \\
        \cmidrule{2-5}
                                      & \multirow{4}{*}{60} & k-means   & 0.383 / 0.381 & 0.225 / 0.219 \\
                                      &                     & CS + SC   & 0.448 / 0.439 & 0.217 / 0.207 \\
                                      &                     & LASSO-SSC & \textbf{0.529} / \textbf{0.484} & \textbf{0.185} / \textbf{0.178} \\
                                      &
                              & OMP-SSC & 0.192 / 0.203 & 0.271 / 0.259 \\
        \midrule
        \multirow{13}{*}{\textbf{C57}} & \multirow{4}{*}{20} & k-means  & 0.345 / 0.338 & 0.189 / 0.181 \\
                                      &                     & CS + SC   & 0.369 / 0.357 & 0.174 / 0.166 \\
                                      &                     & LASSO-SSC & \textbf{0.438} / \textbf{0.391} & \textbf{0.157} / \textbf{0.135} \\
                                      &
                              & OMP-SSC & 0.258 / 0.265 & 0.205 / 0.190 \\
        \cmidrule{2-5}
                                      & \multirow{4}{*}{40} & k-means   & 0.364 / 0.356 & 0.189 / 0.181 \\
                                      &                     & CS + SC   & 0.386 / 0.374 & 0.174 / 0.166 \\
                                      &                     & LASSO-SSC & \textbf{0.453} / \textbf{0.403} & \textbf{0.158} / \textbf{0.138} \\
                                      &
                              & OMP-SSC & 0.208 / 0.217 & 0.203 / 0.212 \\
        \cmidrule{2-5}
                                      & \multirow{4}{*}{60} & k-means   & 0.366 / 0.359 & 0.196 / 0.188 \\
                                      &                     & CS + SC   & 0.398 / 0.387 & 0.183 / 0.174 \\
                                      &                     & LASSO-SSC & \textbf{0.461} / \textbf{0.417} & \textbf{0.161} / \textbf{0.141} \\
                                      &
                              & OMP-SSC & 0.192 / 0.202 & 0.223 / 0.206 \\
        \bottomrule
    \end{tabular}
    \label{tab:results}
\vspace{-0.35cm}
\end{table}

\subsection{Performance measures}
Given the lack of a ground truth, we compare our results both qualitatively and quantitatively. First, we compare the low-dimensional representation of the USVs using t-SNE \cite{tsne}, for the case $K = 40$ clusters. For k-means, to reduce the computational complexity, we first use PCA to reduce the dimensions to 100 before applying t-SNE. For the other methods, we use the embedding produced by spectral clustering as input for t-SNE. Since the outlier data set does not have an embedding, we only plot the inliers data sets.

We also compute the harmonic mean of cosine distance between the centroids of the clusters:
\begin{equation*}
    \bar{d}_{\cos}(\bm{C}) = \left(\frac{1}{K(K-1)} \sum\limits_{i \neq j} \frac{1}{1 - \cos(\bm{c}_i, \bm{c}_j)}\right)^{-1}.
\end{equation*}
Since the harmonic mean is skewed towards small values, a higher harmonic mean of cosine distances can be interpreted as consistently having more diverse centroids. We also compute the standard deviation $\sigma(d_{\cos}(\bm{c}))$ of cosine distances between centroids. Lower standard deviations indicate that centroids are different from each other more consistently.

\begin{figure}[t]
\captionsetup[subfigure]{justification=centering}
\centering
\begin{subfigure}[b]{0.32\columnwidth}
  \centering
  \includegraphics[width=\textwidth]{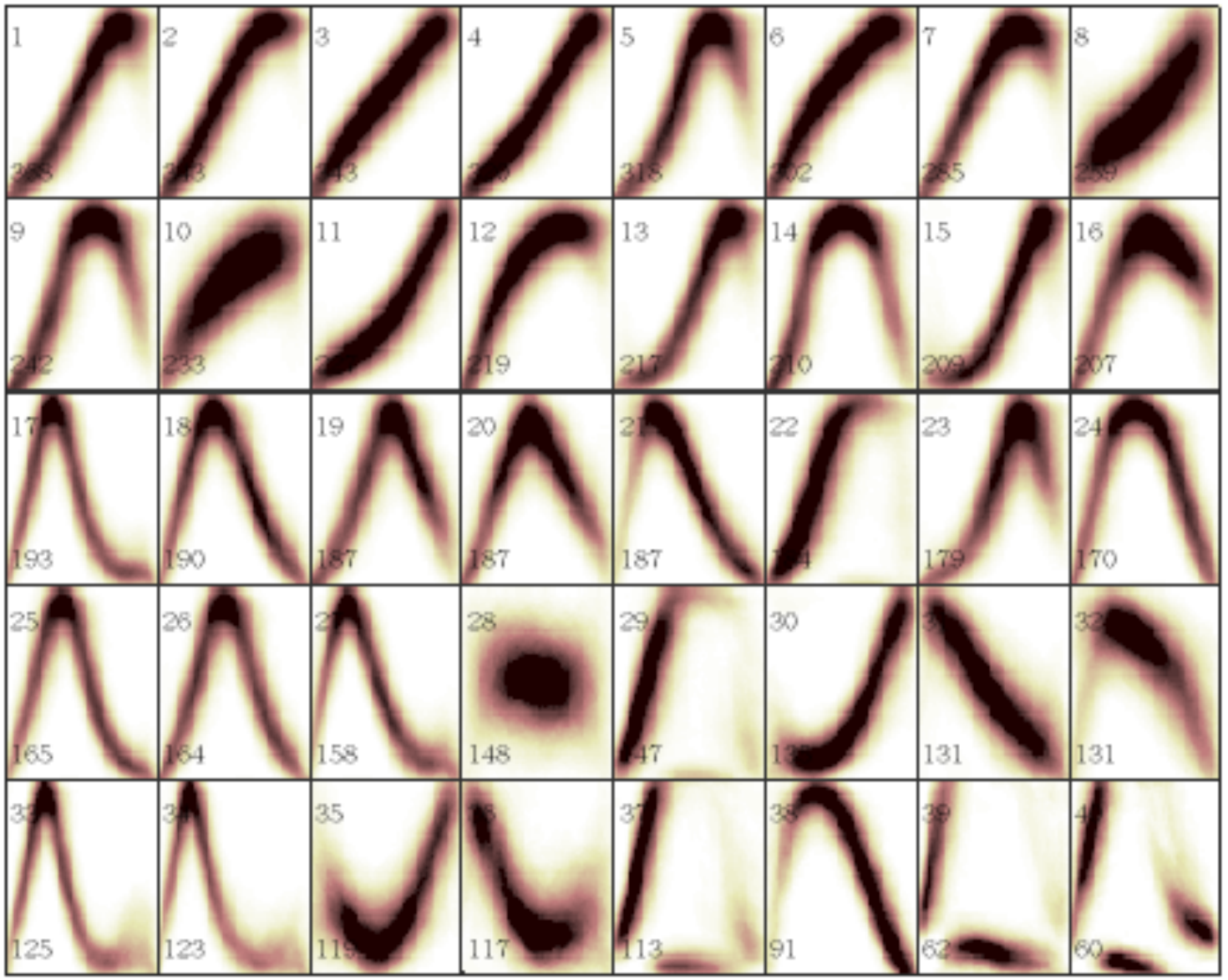}
  \caption{\textbf{DBA}: k-means (baseline)}
\end{subfigure}
\hfill
\begin{subfigure}[b]{0.32\columnwidth}
  \centering
  \includegraphics[width=\textwidth]{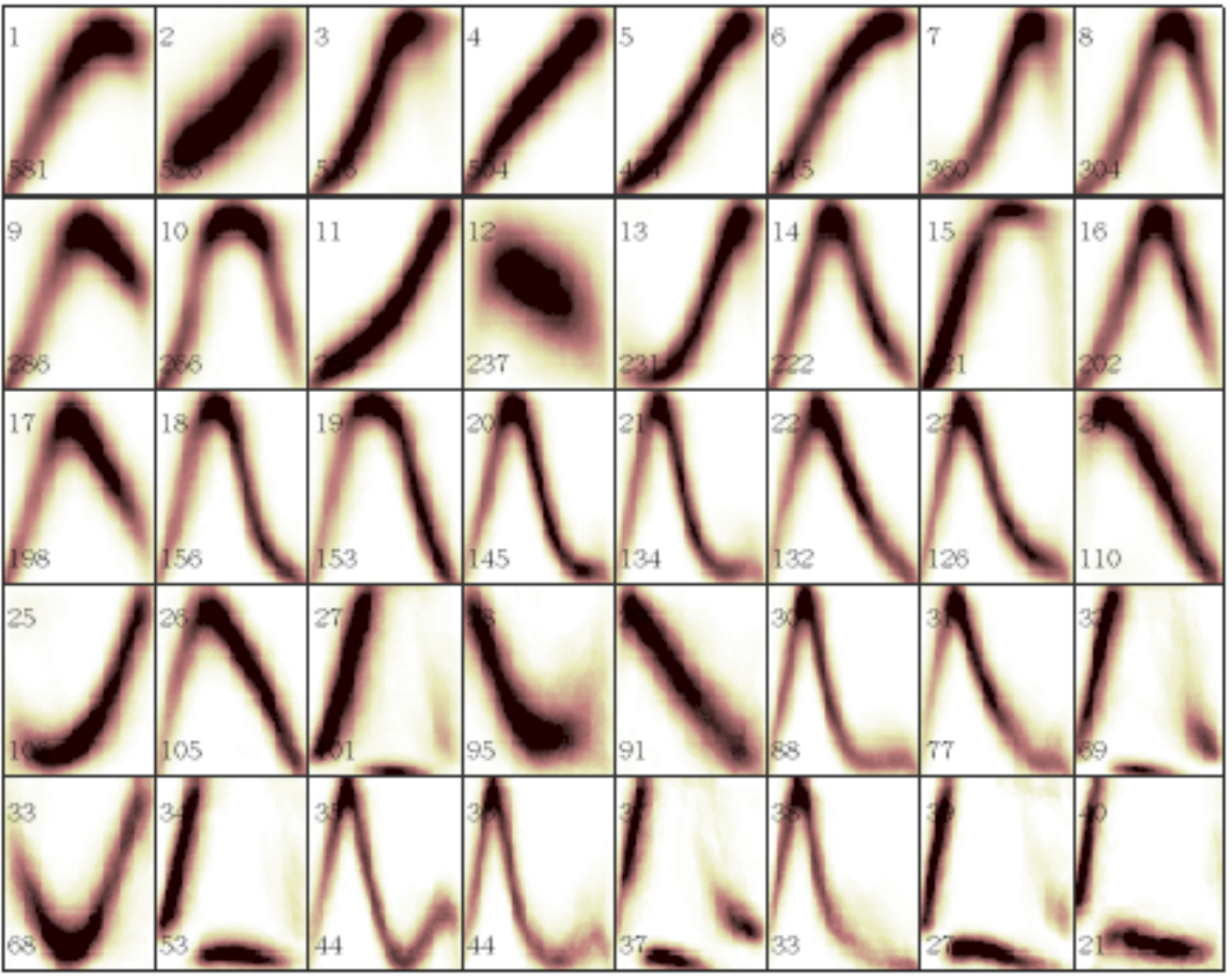}
  \caption{\textbf{DBA}: CS + SC (baseline)}
\end{subfigure}
\hfill
\begin{subfigure}[b]{0.32\columnwidth}
  \centering
  \includegraphics[width=\textwidth]{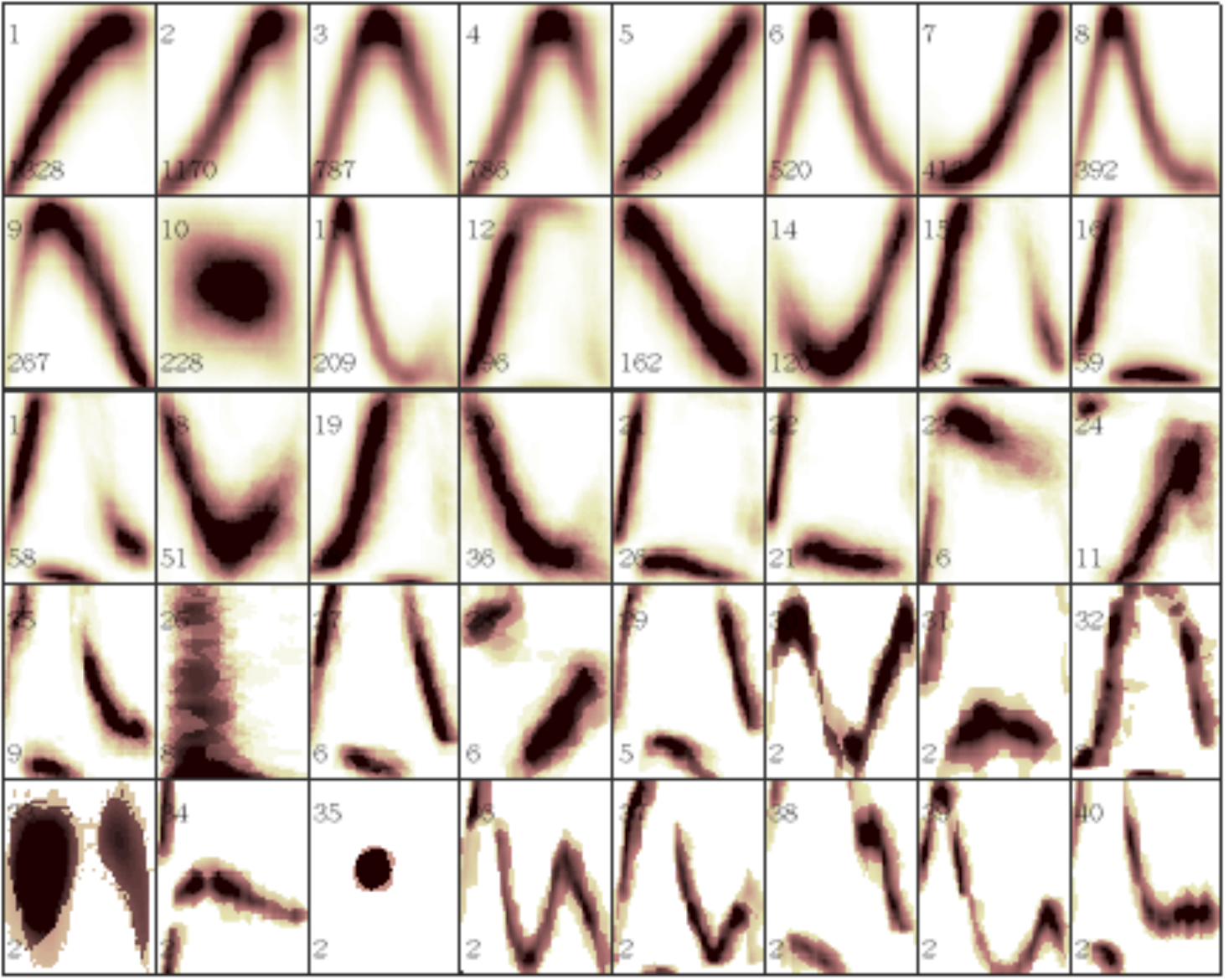}
  \caption{\textbf{DBA}: LASSO- SSC (proposed)}
\end{subfigure}
\begin{subfigure}[b]{0.32\columnwidth}
  \centering
  \includegraphics[width=\textwidth]{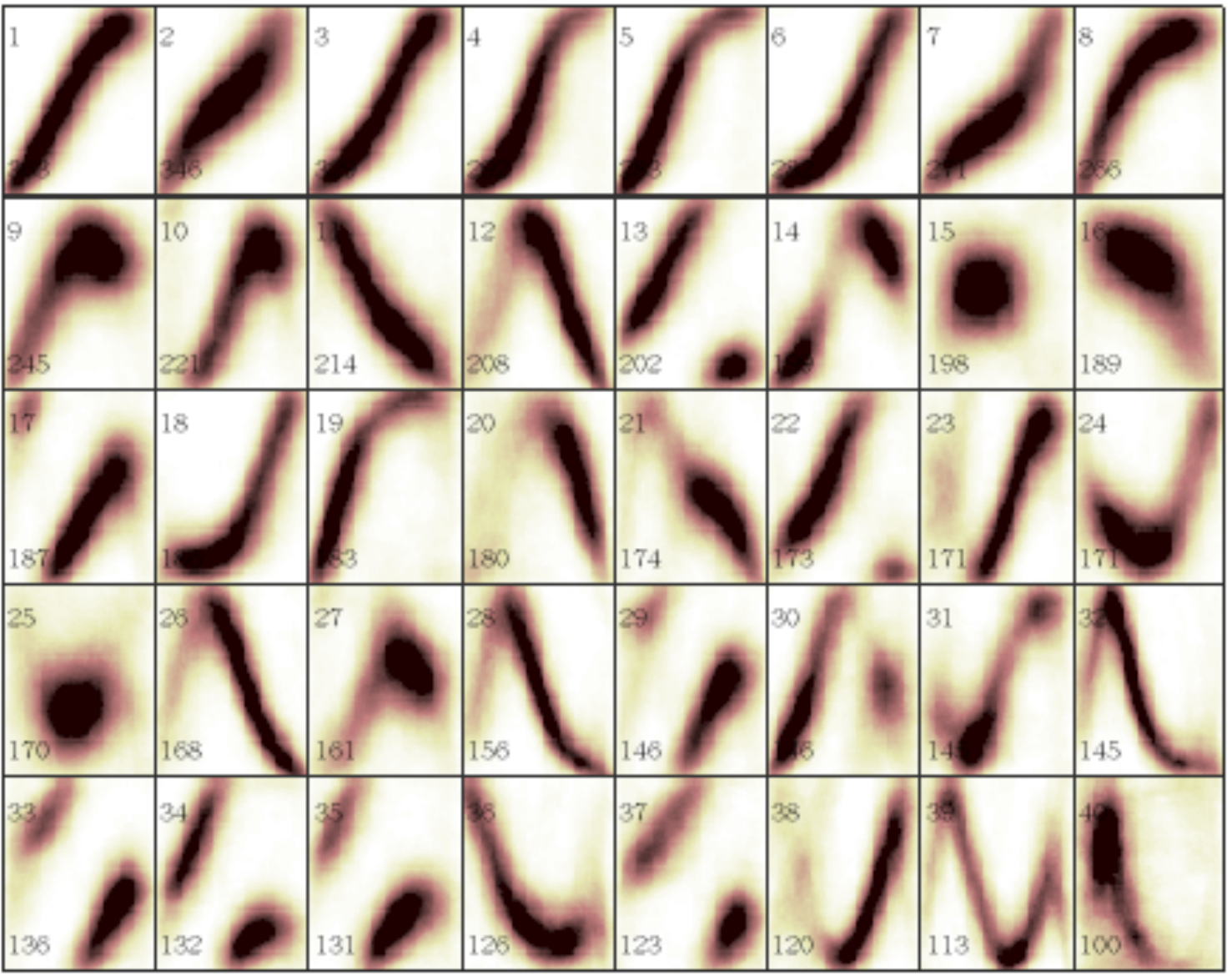}
  \caption{\textbf{C57}: k-means (baseline)}
\end{subfigure}
\hfill
\begin{subfigure}[b]{0.32\columnwidth}
  \centering
  \includegraphics[width=\textwidth]{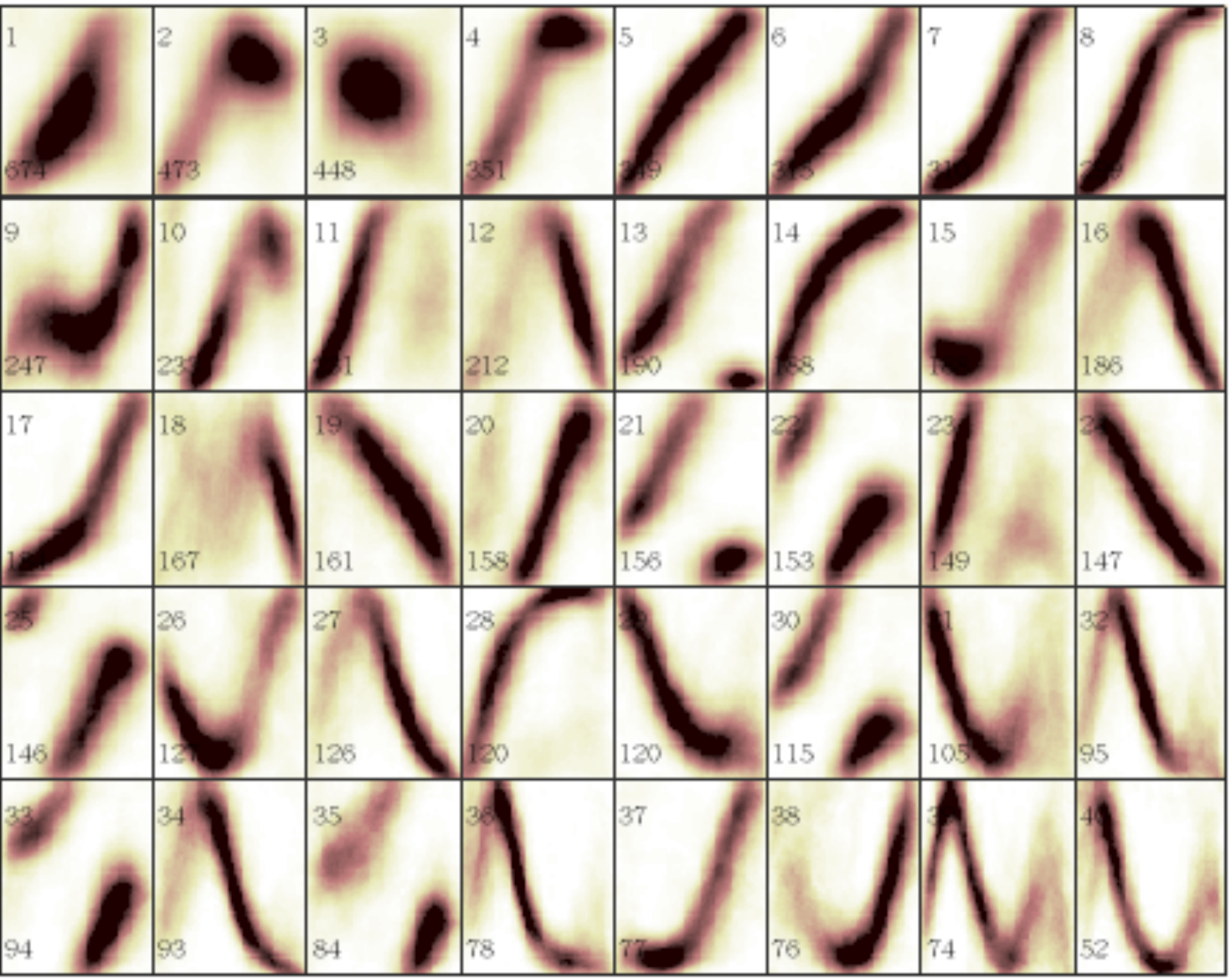}
  \caption{\textbf{C57}: CS + SC (baseline)}
\end{subfigure}
\hfill
\begin{subfigure}[b]{0.32\columnwidth}
  \centering
  \includegraphics[width=\textwidth]{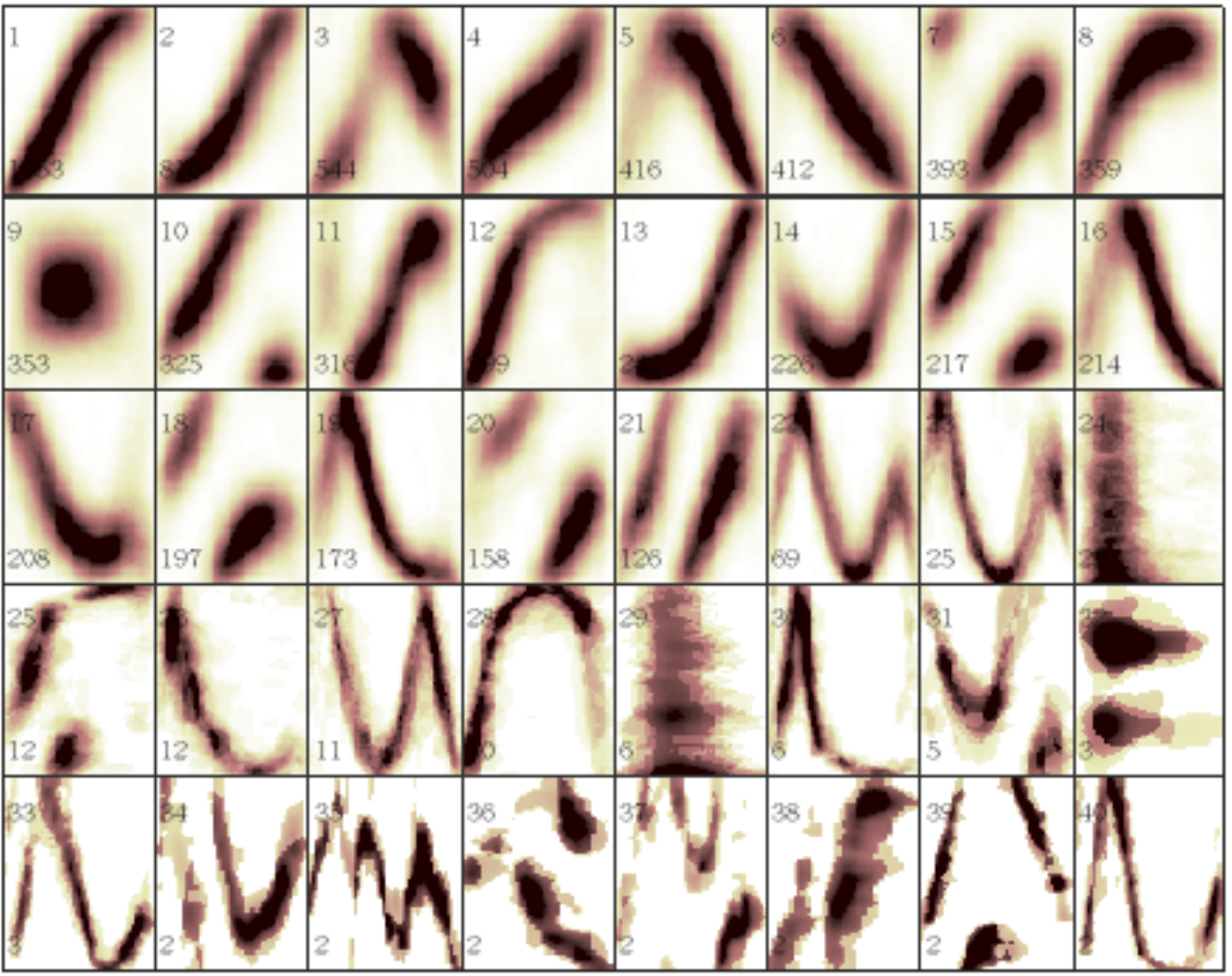}
  \caption{\textbf{C57}: LASSO- SSC (proposed)}
\end{subfigure}
\caption{Centroids of clusters of inliers. In each figure, the top left window contains the cluster with most USVs, and decreases towards the right.}
\label{fig:clustering_centers}
\end{figure}

\section{Results}
We present the quantitative results in \autoref{tab:results}. We observe that the LASSO version of subspace clustering (LASSO-SSC) achieves the best results for all number of clusters $K$ tested in both strains with and without outliers, while the OMP-SSC method yields the worst results across all experiments. When comparing the clustering results on the inlier data set, the best number of clusters is $K = 20$ for DBA. However, for C57 the highest mean cosine distance is achieved for $K = 60$ clusters, while the minimum standard deviation for the cosine distance is achieved at $K = 20$ clusters. When including the outliers, we find that the $\bar{d}_{\cos}(\bm{C})$ of LASSO-SSC decreases more than the other methods. This is because the outliers highly affect several clusters with a small number of samples produced by LASSO-SSC. However, LASSO-SSC achieves the best result for all the cases with outliers. LASSO-SSC has higher $\bar{d}_{\cos}(\bm{C})$ because it tends to cluster similar clusters in other methods into one big cluster.  Therefore, it has less similar centroids, which leads to higher harmonic means of cosine distances between centroids.

We show a qualitative representation in \autoref{fig:clustering_centers}, where the centroids for each cluster are depicted (top left is the cluster containing the largest amount of USVs; the number gets smaller towards the right). k-means and CS + SC have many centroids with similar shapes (for example, several shapes look very similar in the first rows of \autoref{fig:clustering_centers}(a) and (b) and \autoref{fig:clustering_centers}(d) and (e)), while we observe more variability in the centroids for both DBA and C57 using LASSO-SSC. This can be interpreted as our proposed method clustering together the USVs that correspond to a same class or cluster, yielding clusters with less variability of shapes between USVs. Due to the smaller variability within clusters, we see more clusters with noise (by the end of \autoref{fig:clustering_centers}(c,f)), which are clustered together with different shapes by k-means and CS + SC, producing more noisy clusters.

\begin{figure}[t]
\captionsetup[subfigure]{justification=centering}
\centering
\begin{subfigure}[b]{0.3\columnwidth}
  \centering
  \centerline{\includegraphics[width=\textwidth]{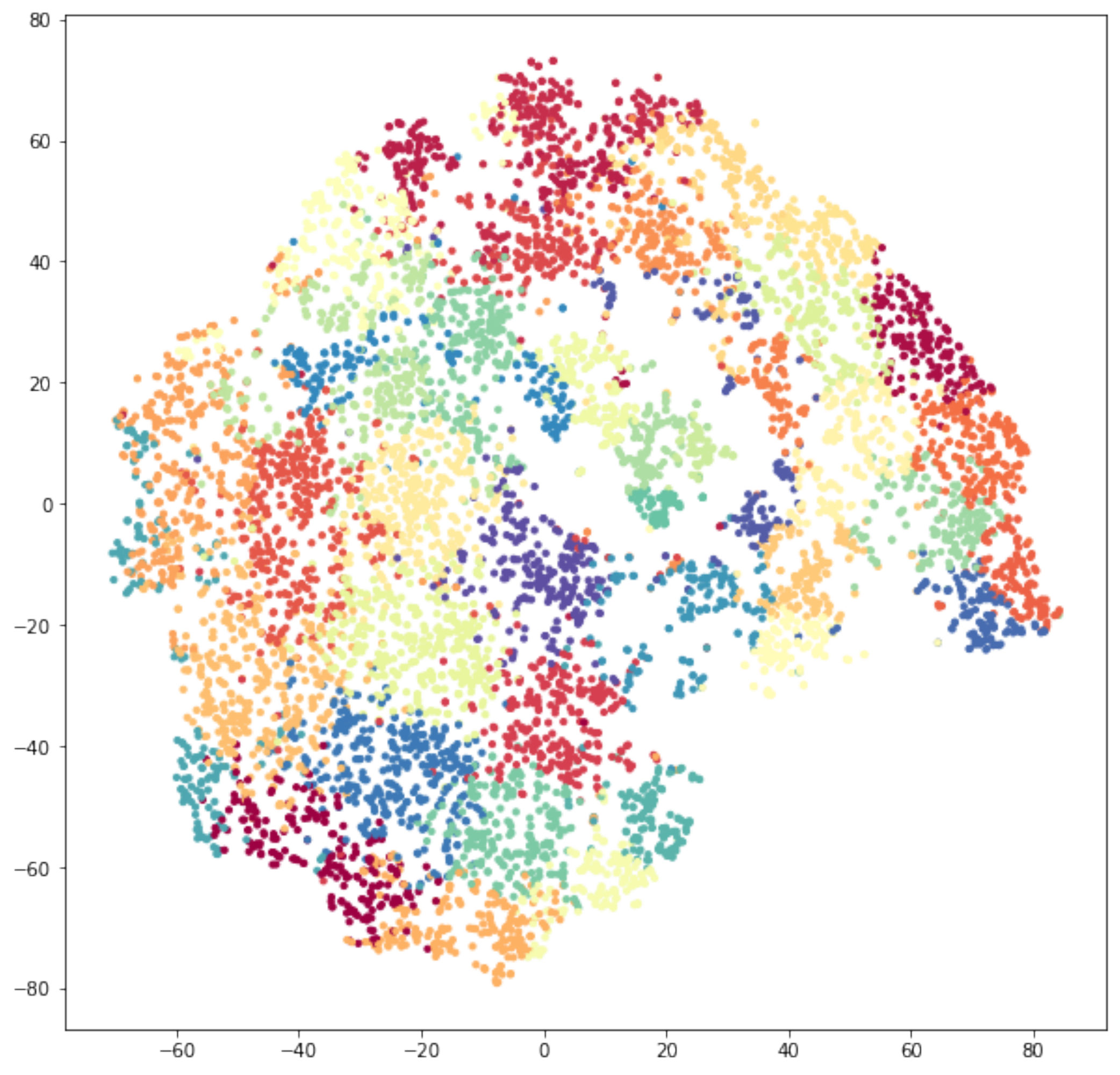}}
  \caption{\textbf{DBA}: k-means (baseline)}
\end{subfigure}
\hfill
\begin{subfigure}[b]{0.3\columnwidth}
  \centering
  \includegraphics[width=\textwidth]{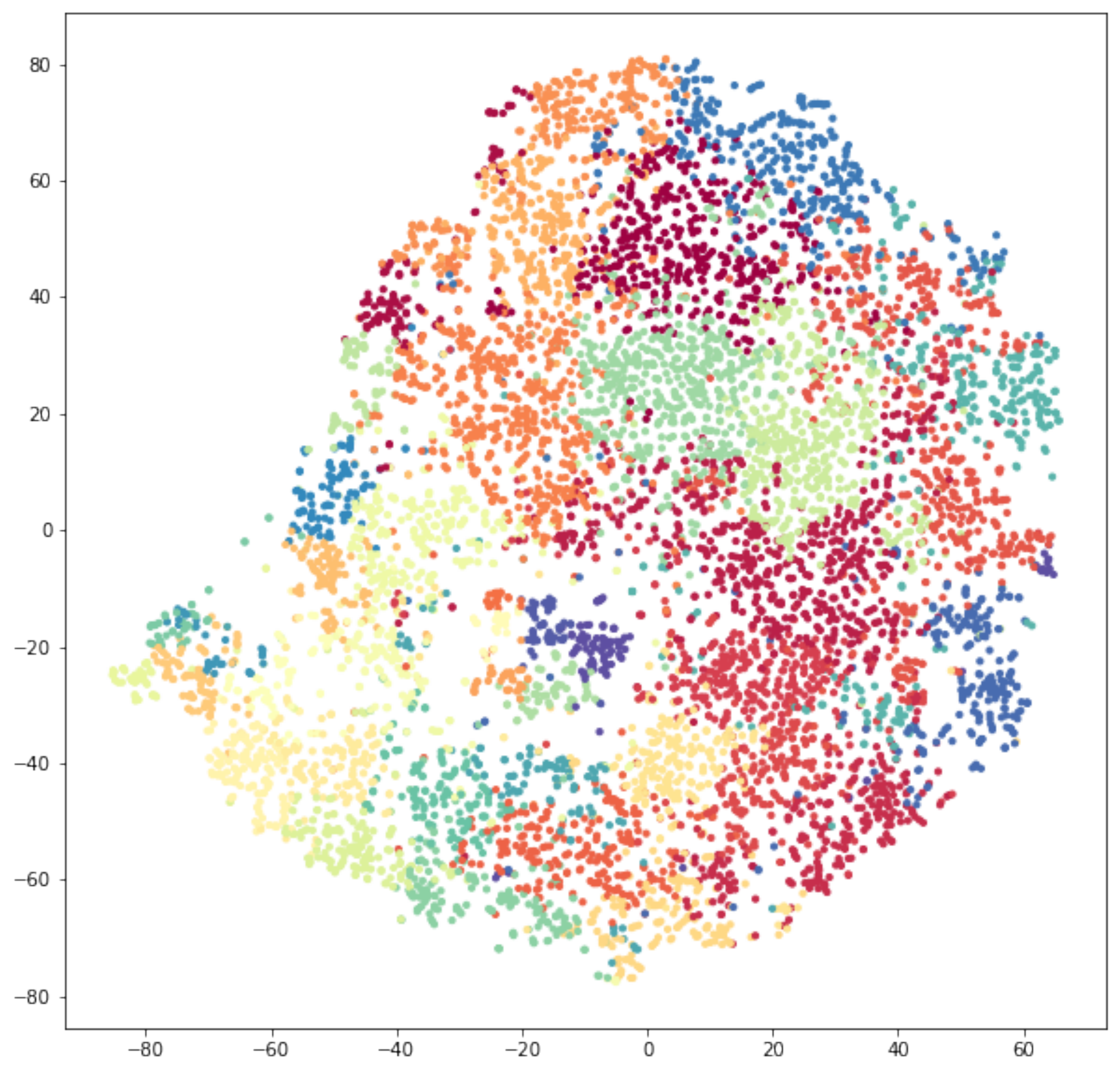}
  \caption{\textbf{DBA}: CS + SC (baseline)}
\end{subfigure}
\hfill
\begin{subfigure}[b]{0.3\columnwidth}
  \centering
  \includegraphics[width=\textwidth]{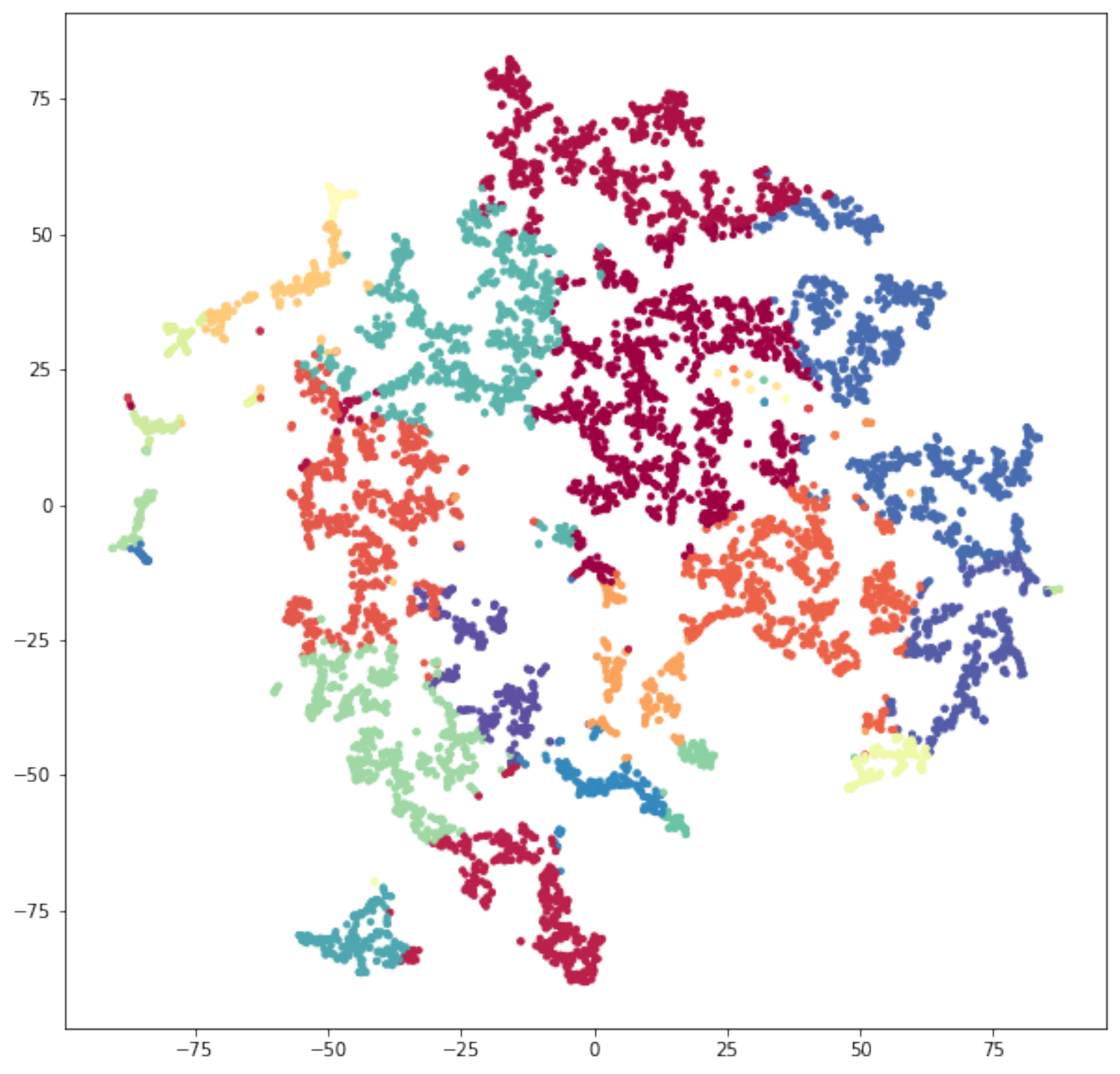}
  \caption{\textbf{DBA}: LASSO-SSC}
\end{subfigure}
\begin{subfigure}[b]{0.3\columnwidth}
  \centering
  \centerline{\includegraphics[width=\textwidth]{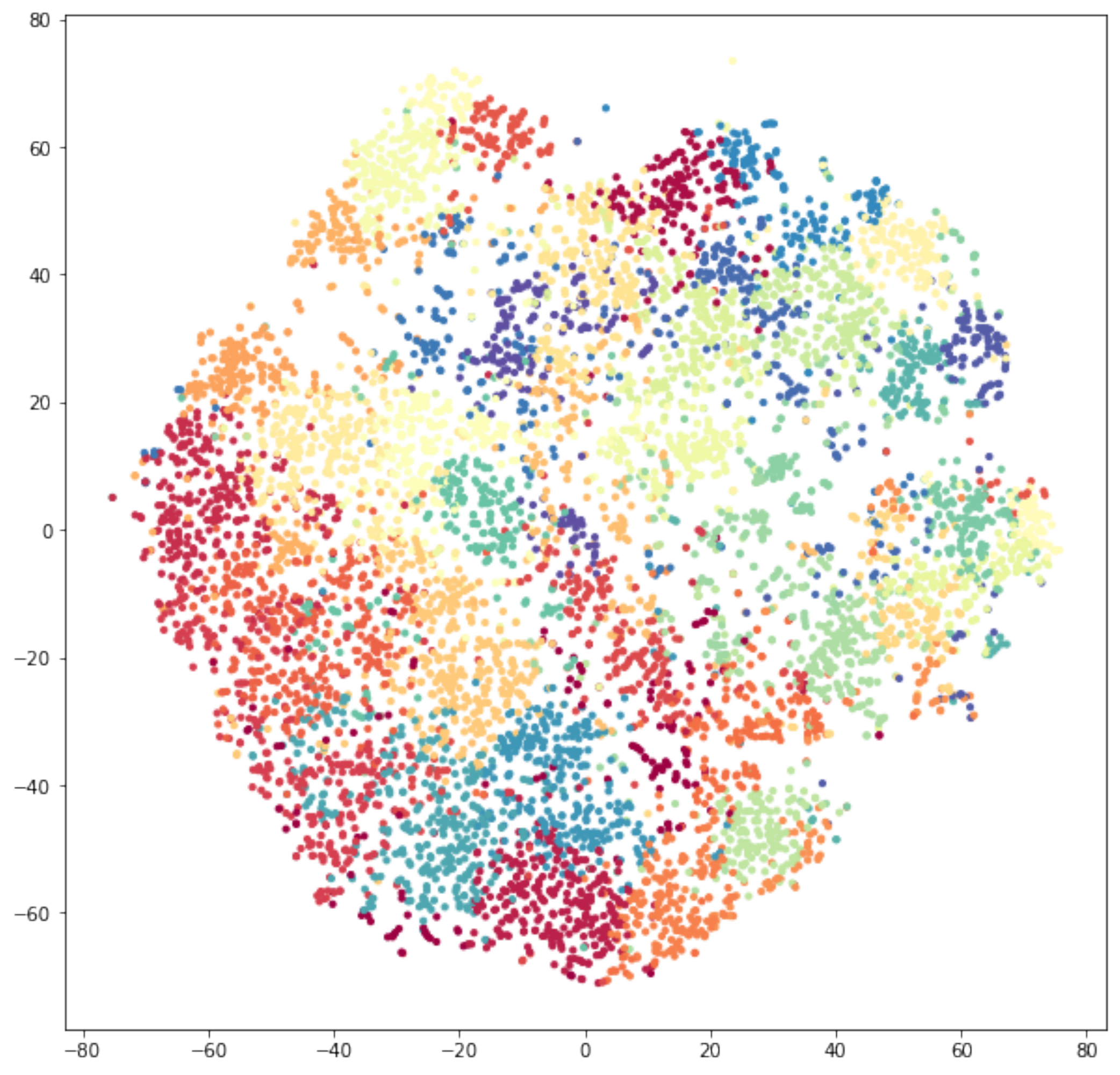}}
  \caption{\textbf{C57}: k-means (baseline)}
\end{subfigure}
\hfill
\begin{subfigure}[b]{0.3\columnwidth}
  \centering
  \includegraphics[width=\textwidth]{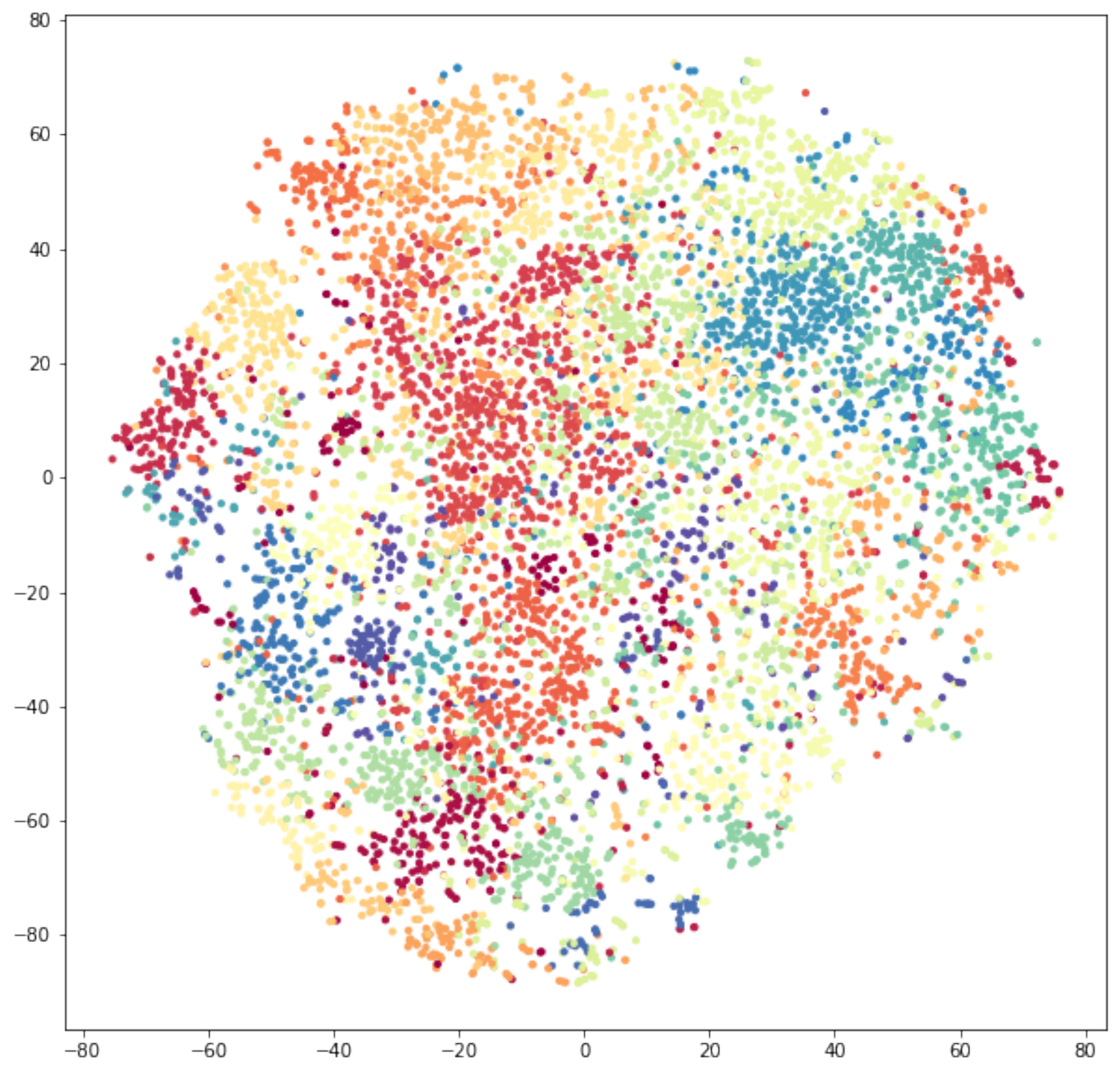}
  \caption{\textbf{C57}: CS + SC (baseline)}
\end{subfigure}
\hfill
\begin{subfigure}[b]{0.3\columnwidth}
  \centering
  \includegraphics[width=\textwidth]{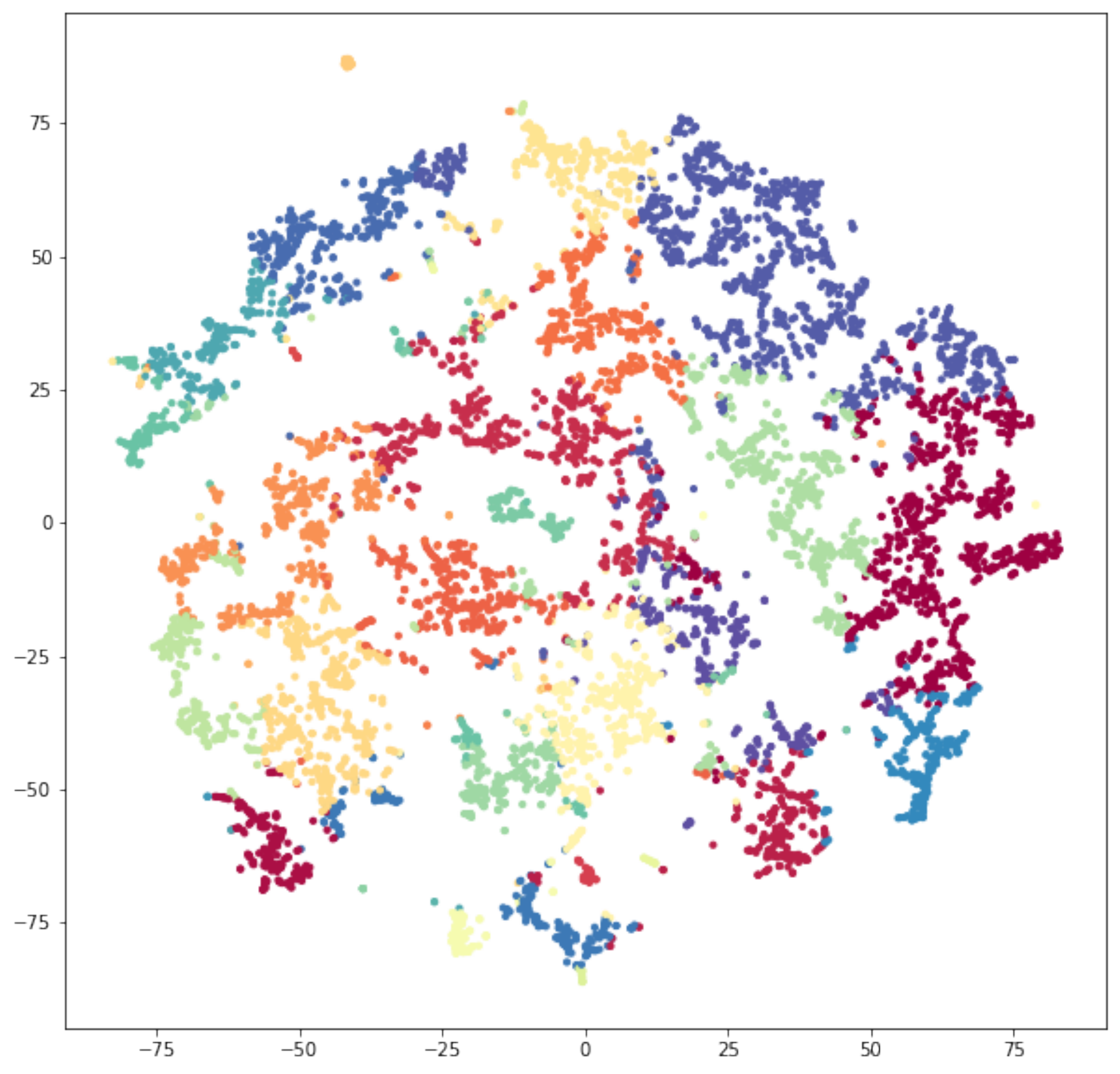}
  \caption{\textbf{C57}: LASSO-SSC}
\end{subfigure}
\caption{t-SNE visualizations of the clusters in 2 dimensions. The embedding computed from the subspace similarity matrix $\bm{A}$ is able to better discriminate among different clusters compared to the cosine similarity between feature vectors.}
\label{fig:tsne_visualization}
\end{figure}

We show different qualitative results in \autoref{fig:tsne_visualization}. These plots show a 2-dimensional representation of the space in which the USVs were clustered, and the colors represent different clusters. We observe that for both strains DBA and C57, the 2-dimensional representations for LASSO-SSC have more distance between clusters than the k-means and CS + SC methods. We omit OMP due to the poor performance shown in \autoref{tab:results}.

\section{Discussion}
One difficulty in clustering task is that some kinds of vocalizations have very few instances, which makes it hard to cluster them into a category. Sparse subspace clustering has a better ability to discover those categories than k-means, which contributes to the diversity of the clusters that can be seen in \autoref{fig:clustering_centers}. Obtaining higher harmonic mean of cosine distances in the proposed method also indicates that subspace clustering leads to clusters with more diversity between each other. These results confirm our original hypothesis that the information in the frequency domain of the USVs is well-modeled as low-dimensional subspaces in a high-dimensional space.

\section{Conclusion}
\label{sec:conclusion}
In this paper, we propose a pipeline to cluster mouse vocalizations based on a two-step approach using sparse subspace clustering. Our method outperforms the previous methods evaluated both through qualitative plots (using t-SNE, \autoref{fig:tsne_visualization}) and our proposed performance measure based on the harmonic mean. Moreover, we find that subspace similarity is a better similarity than cosine similarity to compare USVs. Our approach provides more diverse and cleaner clustering results than previous algorithms.

For future work, we consider several avenues, including: (1) different ways to assess the quality of the clusters, but also looking at the performance of each clustering technique in mice behavior analysis tasks, and (2) including temporal information to aid the clustering.

\section{Acknowledgements}
We thank the USC Viterbi School of Engineering and the Feng Deng Foundation in Tsinghua for their support.


\end{document}